\documentclass[12pt,fleqn]{article}
\usepackage{amsmath}
\usepackage{mathtools}
\usepackage{xcolor,soul}
\usepackage[left]{lineno}
\usepackage{booktabs}
\usepackage{hyperref}

\usepackage[square,numbers,compress]{natbib}

\begin{document}

\begin{center}
\vspace{2cm}     

{\Large\bfseries Critical study of A. Eswari and S. Saravana kumar,
  ``Chronoamperometric response of electrochemical reaction diffusion
  system: a new theoretical and numerical investigation for EC2
  scheme'', in J. Iran. Chem. Soc. \textbf{21}(8), (2024), 2183–2199.}
\vspace{0.8cm}

{\large D. Britz\footnote{britz@chem.au.dk}}\\
  Dept. of Chemistry,Aarhus University, Denmark \\
  ORCID: \url{https://orcid.org/0000-0003-1477-5627} \\[0.4cm]
{\large  J. Strutwolf\footnote{joerg.strutwolf@mb.tu-chemnitz.de}}\\
  Department of Mechanical Engineering \\
  Institute for Print and Media Technology\\
    Chemnitz University of Technology, Chemnitz, Germany\\
    ORCID: \url{https://orcid.org/0000-0002-3881-9770} \\[0.4cm]
\end{center}

\newpage

\noindent\rule[1ex]{\textwidth}{0.5pt}\\[-1.2cm]
\begin{center}
	\subsubsection*{Abstract}
\end{center}
The title paper is discussed critically. There are major
  problems with incorrect statements, irrelevant citations, incorrect
  mathematics leading to incorrect results, which are compared
  with our own simulations.
\noindent\rule[1ex]{\textwidth}{0.5pt}

\subsection*{Introduction}
The paper mentioned in the title shows evidence of careless writing,
meaningless statements, incorrect mathematics and results, as will be
shown below. These results are compared with our own simulation of the
same system under study.

The EC$_2$\;\cite{G.Bard2001} reaction dealt with is rendered as EC2
and in the Theoretical background, it is incorrectly described as
involving two electron transfers. It is in fact a single electron
transfer followed by a homogeneous chemical (dimerisation) reaction.

The electron transfer is intended to be taken as reversible but the ensuing
discussion starts gratuiously with the equations for a quasireversible
reaction, and then goes on to the Nernst relation applying to the reversible
case.

The experiment is said to be cyclic voltammetry (CV) and indeed the
equations rendering the time $T$ dimensionless (clashing with
temperature $T$, also used) uses the scan rate $v$ as is usual for CV,
becoming $t$, which later becomes $\tau$, also serving as the final
time in the results. But clearly the experiment is a jump at $t=0$ to
a fixed potential, which is not CV, so this is constant potential
chronoamperometry. The rate constant $k$ (also called $K$) is made
dimensionless using $v$ and then reverts to $k$. The symbol $\Theta$
refers to a floating potential $E$ (the paper having mentioned
$E-E_0$, used in the discussion of the Nernst equation a few lines
earlier).  If CV were indeed intended, there would have to be a
reversal potential and the scan rate would also have to be normalised,
and these do not appear because this is not CV. The scan rate $v$ then
does not appear again.

The table under the heading ``Theoretical background" has errors
and it would have been better simply to mention that an EC$_2$ reaction
is under discussion.

Most of the Introduction is either trivial, or irrelevant, like the
citations 7-14 and indeed all citations from no. 16 on, which appear to
be padding, most dealing with entirely unrelated subjects.

The geometry of the present electrode is not specified but the
equations are for a planar electrode (not a "planar wire" as mentioned
twice), which is not applicable to a microelectrode, mentioned later,
unless it is at the bottom of a cylindrical hole (Oldham's ``shrouded
electrode''\;\cite{G.Oldh1991b}).

The list of symbols is not complete, lacking $x$, $\Theta$, the symbol
$p$ used briefly later, and the function erf(.); and $c_O$ is said to be a
function of $r$ and $t$, $r$ not being involved in the paper at all,
presumably with a circular disk electrode in mind. It is then dropped
and the spatial dimension is $x$.

The value of $\Theta$ is set in the work to 5, 1 and 0.1. At the value
1 this means the potential is set to the equilibrium potential ($u =
0$). When it is set to 5, this means a positive $u = E - E_0$, meaning
an oxidation, but this does not apply here. Only the case $\Theta =
0.1$ means a jump to a potential where reduction can take place, the
potential now being $u = \log \Theta = -2.3$. So the two sets for
$\Theta=0.1$ are the only part of the Tables that might have produced
valid results. As will be seen, the results nevertheless are
incorrect.

There is more, but these examples suffice here.

\subsection*{Mathematical problems}
Equation (24) reduces to (12), and (25) to (13) except now we have
a new symbol $p$, not defined.

More seriously, and making all that follows, including the tabulated
results and figures, incorrect, is the development of equation (54)
from (52), which assumes that the Laplace transform of a squared term
is the square of its Laplace transform, which is incorrect. Therefore
the solution equations are incorrect, and cannot (and do not)
match either simulated results or (in fact nonexistent) previous
results, as will be shown. Furthermore, the two equations (70) and
(71) for the concentrations as f$(x,\tau,k)$ do not match the
tabulated results nor the real results. Equation (72) for the current
computes a current increasing with time for $t>1$, which is also
incorrect.

The tables have problems. There are numbers for ``This work'',
presumably from the above-mentioned three equations, but they do not
match computed values using the presented equations. There are finite
currents reported for $t=0$, where there is the well known
singularity, also produced by equation (72), so these numbers were not
computed from the solutions. In fact, all current values are wrong. Some
concentration values are fairly close but not equal to values produced
by the authors' equations (70) and (71), but certainly incorrect, as
compared with our own simulations.

Then there are columns headed ``Simulations'', presumably numbers
produced by the MATLAB program shown later. This program does not
work, more on this below. There is a column headed ``Rajendran et al
[15]''.  Reference 15 is to a 2011 paper by Eswari and Rajendran
 which solves the same mechanism as in the paper under
discussion by another method, but in fact reports no numbers. As well, it
deals with the quasireversible case, so the expressions for the
current in that paper are not comparable. The quasireversible case can
be forced close to the reversible case by setting the rate
constants $k_f$ and $k_b$ as used in \cite{Eswa2011b} very high, and
then the currents as calculated from the $\Psi$ function in that paper
for large $t$ are close to but not equal to the column called ``Rajendran
et al'', deviating by about 10\% from these.  So again, the numbers in
these columns need explanation.

The paper's Fig.2 is said, in the legend, to be based on equations
(64) and (65).  The former is a Laplace transform, and the latter is
an inverse Laplace transform resulting in the variable $t$. Neither of
these can produce these concentration profiles, which in fact are
not the result of the reaction mechanism under discussion.

There is a section where equations for a steady state are derived. The
reaction pair dealt with in the paper does not approach a steady
state, except perhaps that the currents go to zero at very long times,
and the concentrations for both $c_O$ and $c_R$ also approach zero at
very long times if the $x$ range is extended to allow this. This
section needs an explanation.

\subsection*{The MATLAB program}
The MATLAB program given in Appendix 1 for the numerically
soliution of the differential equation has some issues that
prevent the execution of the code. The MATLAB PDE solver
\verb|solvepde()|\;\cite{solvepde} is called with an inappropriate set
of parameters (variables and functions). Curiously, the MATLAB
routine \verb|pdepe()|\;\cite{pdepe}, not called in the example,
requires the parameter list as given in the code of the
Appendix. Using \verb|pdepe()| and fixing some other issues (for
example, in line 3 of the code), we were able to run the program
(using MATLAB R2024b and Partial Differential Equation
Toolbox). The program requires the input of a potential, which
does not change with time.  A predefined spatial mesh with
constant intervals of size $\delta x = 1.01$ is used. Such a
coarse mesh will introduce considerable errors in the numerical
approximation of the concentrations, especially adjacent to the
electrode where a reaction layer is formed and the concentration
changes over a small spatial extension. In the program of
Appendix 1, a dimensionless reaction rate of $k=50$ is
defined. The reaction layer thickness can be estimated to be
approximately $1/\sqrt{k} \approx 0.14$. Clearly, the coarse
grid used in the program is not able to adequately reproduce the
concentration within the reaction layer. In the article, Eswari
and Kumar do not provide information about the spatial and
temporal parameter used in their numerical simulations, nor are
convergence tests performed to estimate the quality of the
mesh. Using numerical simulations in this form to demonstrate
the accuracy of their mathematical analysis is therefore
questionable.
\begin{figure}[!b]%
\begin{center}
\includegraphics*[width=0.8\textwidth]{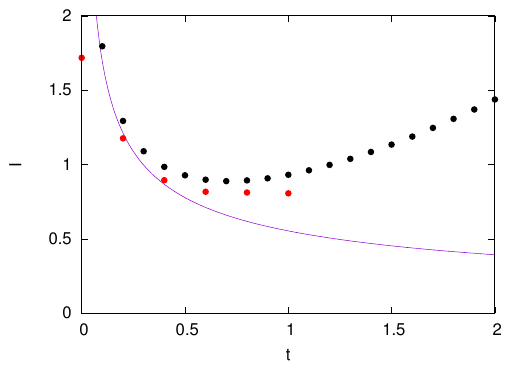}
\caption{Current vs. $t$ for $\Theta=0.1$ and $k=10$. Solid curve,
  the present authors' simulated currents; red points: values reported
  in the paper under discussion as ``This work'' in Table 3; black
  points: currents as computed from equation (72).}\label{fig1}
\end{center}
\end{figure}

\subsection*{Computations}
We have conducted our own simulation of the problem, using the same
dimensionless variables as in the paper under discussion. The method
used was the backward implicit method (BI), linearising the nonlinear
term $c_R'^2$ as decribed in\;\cite[p.164]{db.mono5}. Unequal
intervals in direction $x$ were used, employing the damped exponential
expansion as described\;\cite[p.130]{db.mono5}, the smallest interval
set at $10^{-5}$, and using 100 spatial intervals, and 1000 intervals
in the time direction.  Three-point approximations were used for the
currents. Details on how this is done can be seen in \cite{db.mono5}.

One issue is, how far to go in the $x$ direction, to $x_{lim}$.  As
already mentioned above, it is usual to let $x_{lim} = 6
\sqrt{t_{lim}}$, where $t_{lim}$ is called $\tau$ in the paper under
discussion. This is the time to which a simulation is run. In the
paper $\tau$ is set to 0.1 and 1, but $x_{lim}$ is apparently always
set to unity.  This would result in incorrect values for both current
and concentrations, because then the diffusion range is inadequate and
the far boundary condition at $x_{lim}$ impinges on the
concentrations, being too close.

Fig.\;\ref{fig1} shows currents pertaining to the part of the Tables
for which $\Theta = 0.1$, $k=10$. We have driven the simulation to
$\tau = 2$, using $x_{lim} = 10$. It is seen that the numbers for the
current reported in Table 3 (red points) under ``This work'' are
incorrect, crossing over the correct values at some point. Our
calculation of the authors' equation (71) for $\Psi$, black points,
shows totally incorrect values, increasing for $t>1$. The
concentrations in the other tables are also all incorrect, and an
explanation is needed for the purported small deviations of ``This
work'' from ``Simulations'', given this fact and the fact that the
authors' MATLAB program as presented does not work.

\section*{Conclusion}\label{Concl}
In conclusion, the paper by Eswari and ``kumar'' consists of
meaningless text, incorrect mathematics and, inevitably, incorrect
results, comparing with nonexistent data and an unworkable MATLAB
program that cannot have produced the ``simulated'' results. The paper
should be retracted, unless the authors can give cause, addressing the
above mentioned issues, why it should not be.

\newpage

\begin{thebibliography}{6}
\providecommand{\natexlab}[1]{#1}
\providecommand{\url}[1]{\texttt{#1}}
\expandafter\ifx\csname urlstyle\endcsname\relax
  \providecommand{\doi}[1]{doi: #1}\else
  \providecommand{\doi}{doi: \begingroup \urlstyle{rm}\Url}\fi

\bibitem[Bard and Faulkner(2001)]{G.Bard2001}
A.~J. Bard and L.~R. Faulkner.
\newblock \emph{{Electrochemical Methods 2nd Ed.}}
\newblock John Wiley, New York, 2001.

\bibitem[Oldham(1991)]{G.Oldh1991b}
K.~B. Oldham.
\newblock {The short-time chronoamperometric behaviour of an electrode of
  arbitrary shape}.
\newblock \emph{J. Electroanal. Chem.}, 297:\penalty0 317--348, 1991.

\bibitem[Eswari and Rajendran(2011)]{Eswa2011b}
A.~Eswari and L.~Rajendran.
\newblock {Mathematical Modeling of Cyclic Voltammetry for EC$_2$ Reaction}.
\newblock \emph{Russ. J. Electrochem.}, 47:\penalty0 191--199, 2011.

\bibitem[sol()]{solvepde}
\verb|http://www.mathworks/help/pde/ug/pde.pdemodel.solvepde.html|.
\newblock Link checked 25/09/25.

\bibitem[pde()]{pdepe}
\verb|http://www.mathworks/help/pde/ug/pde.pdemodel.solvepde.html|.
\newblock Link checked 25/09/25.

\bibitem[Britz and Strutwolf(2016)]{db.mono5}
D.~Britz and J.~Strutwolf.
\newblock \emph{Digital Simulation in Electrochemistry, 4th Ed.}
\newblock Springer, Berlin, 2016.
\newblock ISBN 978-3-319-30292-8, 978-3-319-30290-4.

\end{thebibliography}

\end{document}